\documentstyle[preprint,prd,aps]{revtex}
\begin{document}
\draft\tightenlines
\title{How to measure the mass of the $W$}
\author{Matthew H. Austern}
\address{Lawrence Berkeley Laboratory, University of California,
         Berkeley, California, 94720 \\ and \\
         Physics Department, University of California at Berkeley,
         Berkeley, California, 94720
}
\author{Robert N. Cahn}
\address{Lawrence Berkeley Laboratory, University of California,
         Berkeley, California, 94720
}
\date{\today}
\preprint{LBL-33780}
\maketitle
\begin{abstract}
We perform a numerical calculation of the total cross section
$\sigma(e^+e^- \rightarrow W^+W^-)$ as a function of energy, taking
into account the finite width of the $W$ and the most important
radiative corrections.  We present these results, in tabular form, for
several values of $M_W$.  Using these results, we investigate running
strategies for integrated luminosities that might be available at
LEP~200 and estimate the accuracy to which it will be possible to
determine the mass of the $W$ by measuring this cross section near
threshold.  With an integrated luminosity of $100\ {\rm pb}^{-1}$ it
should be possible to achieve a precision of $100\ {\rm MeV}$, and
with an integrated luminosity of $500\ {\rm pb}^{-1}$, a precision of
$60\ {\rm MeV}$.

\end{abstract}


\narrowtext
\section{Introduction}
The gauge sector of the Standard Model of electroweak physics is
described by three parameters, which can be chosen to be $\alpha$, the
electromagnetic coupling constant, and $M_Z$ and $M_W$, the masses of
the weak gauge bosons.  Of these three, $M_W$ is by far the least well
known.  A more practical choice of parameters is $\alpha$, $M_Z$, and
$G_F$, where $G_F = (1.16639 \pm 0.00002) \times 10^{-5}\ {\rm
GeV}^{-2}$ is the Fermi decay constant.  The mass of the $W$, then,
can be predicted in terms of these three parameters; this prediction
depends on some additional physics as well, most notably the masses of
the top quark and the Higgs boson.  Despite this dependence, and
despite the large uncertainty in the measured value of $M_W$, a
comparison of $M_W$ to the predicted value is already a stringent test
of the Standard Model.

The $W$ mass is currently measured~\cite{pdg} to be $80.22 \pm 0.26\
{\rm GeV}$, where this value is obtained by direct reconstruction of
$W$ events at $\bar{p}p$ colliders; it is expected~\cite{CDF-future}
that future experiments at the Fermilab Tevatron will be able to
reduce this uncertainty to less than $150\ {\rm MeV}$.  This is still,
however, a much larger relative uncertainty than in the other
fundamental parameters of the Standard Model.

One of the major motivations for upgrading LEP to an energy of
$\sqrt{s} \approx 200\ {\rm GeV}$ is to measure $M_W$ more precisely.
Several methods have been proposed, including measurement of the end
point of the leptonic decay spectrum, direct reconstruction of the
invariant mass of the $W$'s decay products~\cite{beam-line}, and
measurement of the threshold for $W^+W^-$ pair production.  We will
discuss the last of these methods.

In principle, $M_W$ can be measured simply by determining the end
point of the $W^+W^-$ spectrum; in practice, however, there is no
sharp end point, since the $W$ has a finite width, with $\Gamma_W =
2.12 \pm 0.11\ {\rm GeV}$.  Initial-state radiation results in a
further smearing of the cross section.  In the regime $\sqrt{s}
\approx 2 M_W$, which we will call the threshold region, the cross
section rises steeply as a function of energy, varying from about $2\
{\rm pb}$ at $\sqrt{s} = 160\ {\rm GeV}$ to $4\ {\rm pb}$ at $\sqrt{s}
= 162\ {\rm GeV}$. A precise measurement of the cross section in this
region provides essentially the same information as would be obtained
from determination of the end point, if a sharp end point existed.

Once the finite $W$ width and initial-state radiation are taken into
account, there is no simple analytic form for the $W^+W^-$ production
cross section.  Accordingly, we have calculated this cross section
numerically, and used these results to estimate the precision to which
it will be possible to measure $M_W$ at LEP~200.
        
\section{Calculation of the $W$ pair production cross section}
At tree level, $W^+W^-$ production at an $e^+e^-$ collider proceeds by
the diagrams in Fig.~\ref{production-diagrams}.  Evaluation of these
diagrams is straightforward; unfortunately, if done na{\"\i}vely, it
yields an unrealistic answer near threshold.  The $W$ decays rapidly,
and it is incorrect to put it on-shell in the final state.

A more realistic answer~\cite{cross-section-paper} is given by
\begin{equation}                \label{double-smear}
\sigma(s) = \int_0^s \! dm_1^2 \, \rho(m_1^2)
            \int_0^{(\sqrt{s} - m_1)^2} \!\! dm_2^2 \, \rho(m_2^2) \,
            \sigma_0 (s, m_1, m_2),
\end{equation}
where $\sigma_0$ is the cross section for production of two off-shell
$W$'s with masses $m_1$ and $m_2$, and where $\rho$ is a weight
factor, 
\begin{equation}
\label{weight-factor}
\rho(m^2) = {1\over\pi}  {\Gamma_W \over M_W}
            {m^2  \over  (m^2 - M_W^2)^2 + m^4 \Gamma_W^2 / M_W^2}.
\end{equation}
This expression for $\rho$ differs slightly from that given in
Ref.~\cite{cross-section-paper}, but the difference is numerically
inconsequential.  A derivation is given in the Appendix.

The cross section for pair-production of off-shell $W$'s has been
calculated by Muta, Najima, and Wakaizumi~\cite{cross-section-paper},
and their result is
\widetext
\begin{equation}
\sigma_0(s, m_1, m_2) = 
 {1 \over 32 \pi s^2 m_1^2 m_2^2}
  \left( a_{\gamma\gamma} + a_{ZZ} + a_{\gamma Z} +
         a_{\nu\nu} + a_{\nu Z} + a_{\nu\gamma}  \right),
\end{equation}
where
\begin{mathletters}
\begin{eqnarray}
a_{\gamma\gamma} &=& {e^4 \over s^2} G_1(s, m_1, m_2),          \\
a_{ZZ} &=& {g^4 \over 16} 
           {(1 - 4\sin^2\theta_W)^2 + 1         \over 
                (s - M_Z^2)^2 + M_Z^2 \Gamma_Z^2
           }
           G_1(s, m_1, m_2),            \\
a_{\gamma Z} &=& {e^2 g^2 \over 2}
                 {1 - 4\sin^2\theta_W \over s} 
                 {s - M_Z^2 \over (s - M_Z^2)^2 + M_Z^2 \Gamma_Z^2}
                 G_1(s, m_1, m_2),              \\
a_{\nu\nu} &=& {g^4\over 8} G_2(s, m_1, m_2),           \\
a_{\nu Z} &=& {g^4 \over 8}
              (2 - 4\sin^2\theta_W) 
              {s - M_Z^2 \over (s - M_Z^2)^2 + M_Z^2 \Gamma_Z^2}
              G_3(s, m_1, m_2),                         \\
a_{\nu\gamma} &=& {e^2 g^2 \over 2} {1\over s}
                  G_3(s, m_1, m_2),
\end{eqnarray}
\end{mathletters}
\begin{mathletters}
\begin{eqnarray}
G_1 &=& \:\:\, \lambda^{3/2}
    \left[{\lambda\over6} + 2(s(m_1^2 + m_2^2) + m_1^2 m_2^2)
    \right],                                            \\
G_2 &=& \:\:\, \lambda^{1/2}
    \left[{\lambda\over6} + 2(s(m_1^2 + m_2^2) - 4 m_1^2 m_2^2)
    \right]
  + 4 m_1^2 m_2^2 (s - m_1^2 - m_2^2) F,                \\
G_3 &=& - \lambda^{1/2}
   \left[{\lambda\over6} (s + 11 m_1^2 + 11 m_2^2)
         + 2s (m_1^4 + 3 m_1^2 m_2^2 + s_2^4) 
         - 2(m_1^6 + m_2^6) \right]                     \\ \nonumber
   && \quad\quad \mbox{}+ 
	4m_1^2 m_2^2 \bigl(s(m_1^2 + m_2^2) + m_1^2 m_2^2\bigr) F,
\end{eqnarray}
\end{mathletters}
\begin{equation}
\lambda = s^2 + m_1^4 + m_2^4 
                - 2 (s m_1^2 + s m_2^2 + m_1^2 m_2^2),
\end{equation}
and
\begin{equation}
F = \ln\left({s - m_1^2 - m_2^2 - \sqrt\lambda          \over
              s - m_1^2 - m_2^2 + \sqrt\lambda} \right).
\end{equation}
\narrowtext

When both $W$'s are on-shell, {\it i.e.}, $m_1 = m_2 = M_W$, this
reduces to the well-known form~\cite{on-shell-cross-section}
\widetext
\begin{eqnarray}
\sigma(s) = {\pi\alpha^2\beta \over 2 s \sin^4\theta_W} \Bigg[&&
\left( 1 + 2 {M_W^2 \over s} + 2 {M_W^4 \over s^2} \right)
{L\over\beta}
- {5\over4}                     
\\              \nonumber
&& {}+
{M_Z^2 (1 - 2\sin^2\theta_W) \over s - M_Z^2}
\left( 2 {M_W^4 \over s^2} \left(1 + 2 {s \over M_W^2} \right)
                           {L\over\beta} 
      - {1\over12}{s\over M_W^2} - {5\over3} - {M_W^2 \over s} \right)
\\                      \nonumber
&& {}+
{M_Z^4 (1 - 2\sin^2\theta_W)^2 \over 48 (s - M_Z^2)^2}
\beta^2
\left( {s^2 \over M_W^4} + 20 {s \over M_W^2} + 12 \right) \Bigg],
\end{eqnarray}
\narrowtext\noindent
where $\beta$ is the speed of the $W$'s in the center of mass frame
and
\begin{equation}
L = \ln {1+\beta\over1-\beta}.
\end{equation}

Numerical evaluation of Eq.~\ref{double-smear} is somewhat
time-consuming, since it involves two nested integrals, both of which
are over sharp peaks.  Fortunately, it is possible to make a vastly
simplifying approximation:
\begin{equation}                \label{single-smear}
\sigma(s) \approx \int_0^{(\sqrt{s} - M_W)^2} dm^2 \tilde\rho(m^2) \,
                  \sigma_0 (s, m, M_W),
\end{equation}
where the weight factor $\tilde\rho$ is exactly the same as the weight
factor $\rho$ defined in Eq.~(\ref{weight-factor}), except that the
$W$ width, $\Gamma_W$, is replaced by $2\Gamma_W$.

Fig.~\ref{single-smear-comparison} compares the cross section
given by Eq.~(\ref{double-smear}) to that given by
Eq.~(\ref{single-smear}).  The approximation of
Eq.~(\ref{single-smear}), of course, breaks down completely near 
$\sqrt{s} \lesssim M_W$.  This, however, is not the domain of
interest, and in the region near threshold, $\sqrt{s} \approx 2M_W$,
the approximation is excellent.

Several other higher-order effects are also significant.  The most
important are the running values of the gauge coupling constants, and
initial-state radiation.  The first of these can be taken into account
simply by using the values for the coupling constants renormalized at
a scale near $M_W$; we use the values measured~\cite{amaldi-paper} at
$M_Z$.  Inclusion of initial-state radiation requires additional work.

Although initial-state radiation is a purely electromagnetic effect,
and is thus suppressed by a factor of $\alpha$, it is nonetheless
significant because it is enhanced by a factor of $\ln(M_W^2/m_e^2)$,
representing the presence of two very different energy scales.  Using
the formalism of Kuraev and Fadin~\cite{initial-state-paper}, 
it is possible to sum all orders of initial-state radiation by
performing a single integral:
\widetext
\begin{equation}        \label{initial-state-equation}
\sigma(s) = t \int_0^{\sqrt{s}/2}  dk
\left[   {1 \over k}
         \left(1 + {3t\over4}\right) 
         \left( {2k \over \sqrt{s}} \right)^t
      -  {2 \over \sqrt{s}} 
         \left( 1 - {k\over\sqrt{s}} \right)
\right]
\sigma_0 \left[ \left( \sqrt{s} - k \right)^2 \right],
\end{equation}
\narrowtext\noindent
where
\begin{equation}
t = {2 \alpha \over \pi} 
    \left(  \ln\left( {M_W^2 \over m_e^2} \right) - 1 \right)
  \approx 0.1065.
\end{equation}
The second term in the integral represents single-photon hard
bremsstrahlung, while the first is the result of summing all orders of
soft photon emission.  This formalism has previously been used to
include the effects of initial-state radiation to all orders in the
calculation of $\sigma(e^+e^-) \rightarrow Z$~\cite{cahn}. The
difference between this result and the ${\cal O}(\alpha)$ calculation
can be substantial.

Several groups~\cite{one-loop-calculations} have performed full
one-loop calculations of $\sigma(e^+e^- \rightarrow W^+W^-)$.  Our
calculation is considerably simpler, but contains most of the relevant
physics, including initial-state radiation beyond leading order in
$\alpha$.

Fig.~\ref{cross-section-plot} shows the results of including
initial-state radiation in the calculation of $\sigma(e^+e^-
\rightarrow W^+W^-)$.  Note that initial-state radiation makes a
contribution roughly equal in magnitude to that of the $W$'s width,
and that it has the effect of making the threshold for $W$ pair
production even less sharp.

Although there is no longer a sharp threshold, $\sigma(e^+e^-
\rightarrow W^+W^-)$ still depends strongly on the value of the $W$
mass; as shown in Fig.~\ref{several-masses}, this dependence is
strongest for $E \approx M_W$.  It is possible, then, to measure $M_W$
by studying the threshold behavior of $\sigma(W^+W^-)$.
Table~\ref{mass-dependence-table} presents the same information, for
masses within two standard deviations of the current central value, in
tabular form.

Note that the behavior of $\sigma$ in the threshold region is mainly
due to kinematic effects~\cite{model-dependence-paper}, so it is
proper to use it for a determination of $M_W$.  The behavior of the
cross section near the peak, at around $\sqrt{s} \approx 220\ {\rm
GeV}$, is also of interest, but for different reasons.  At this point,
the total cross section depends on delicate gauge cancellations, so it
is a sensitive probe of the Standard Model gauge structure.

\section{Measurement of $M_W$ at LEP 200}
\subsection{Lower bound on the statistical error}
                \label{statistical-section} 
The most favorable situation possible would be if there were no
uncertainty in the luminosity or the energy of the beam, and if there
were no theoretical uncertainties; in this case, as seen in
Fig.~\ref{several-masses}, measurement of the cross section at even a
single point could determine $M_W$, and the error in $M_W$ would be
purely statistical.

The error in the measured value of $M_W$ is given by
\begin{equation}
\delta M = \left|{d\sigma \over dM}\right|^{-1} \delta\sigma,
\end{equation}
and, if the error is assumed to be purely statistical, 
\begin{equation}
\delta\sigma =  {\delta N \over {\cal I}},
\end{equation}
where ${\cal I} = \int d {\cal L}$ is the integrated luminosity and
$\delta N$ is the statistical error in the number of $W^+W^-$ events
observed.  For $N \gg 1$ the statistical error approaches $\sqrt{N}$,
and
\begin{equation}
\delta\sigma = \sqrt{\sigma\over{\cal I}}.
\end{equation}

The quantity $d\sigma/dM$ can be read off from
Table~\ref{mass-dependence-table}, but there is a simpler way to
obtain a rough estimate.  Near threshold, as seen in
Fig.~\ref{several-masses}, the most important effect of changing $M_W$
is simply a shift in the $W^+W^-$ spectrum, {\it i.e.},
\begin{equation}
\sigma(E, M_W + \delta M) \approx 
\sigma(E-\delta M, M_W),
\end{equation}
where $E = \sqrt{s}/2$.
Roughly, then, 
\begin{equation}
\left|{\partial\sigma\over\partial M}\right| \approx
\left|{\partial\sigma\over\partial E}\right|,
\end{equation}
and, measuring the cross section at some particular value of $E$, the
error in $M_W$ is given by
\begin{equation}
\delta M_W \approx
\sqrt{\sigma} \left| {\partial\sigma\over\partial E} \right|^{-1}
              {1\over\sqrt{{\cal I}}}.
\end{equation}

Independent of ${\cal I}$, then, the optimum energy at which to make
this measurement is where $|\partial\sigma/\partial E| / \sqrt\sigma$
is maximized; this is at $E \approx M_W$.  The precision that may be
attained is roughly
\begin{equation}
\delta M_W \approx
       {870\ {\rm MeV}/{\rm pb}^{1/2}   \over  \sqrt{{\cal I}}}.
\end{equation}
With an integrated luminosity of $1000\ {\rm pb}^{-1}$, this would
mean $\delta M_W \approx 30\ {\rm MeV}$, which is comparable to the
precision to which $M_Z$ is known.

As a rough guide to systematic errors, we note that if the measured
value of $\sigma$ depends multiplicatively on some parameter $C$, then
\begin{equation}
\delta M_W =
        \sigma
        \left| {\partial\sigma \over \partial M} \right| ^ {-1}
        {\delta C \over C}.
\end{equation}
For example, the measured cross section depends multiplicatively on
the detector efficiency, and on the luminosity.  Certain sources of
theoretical error can also, at least approximately, be represented
this way.  Using the values found in
Table~\ref{mass-dependence-table}, we can rewrite this relation as
\begin{equation}
\delta M_W \approx 1.7\ {\rm GeV} {\delta C \over C}.     
			\label{norm-error-equation}
\end{equation}

\subsection{Theoretical error}          \label{theoretical-error}
There are three potential sources of theoretical error in this
measurement: Model dependence, uncertainty in the input parameters,
and incomplete inclusion of radiative corrections.

In general, $\sigma(e^+e^- \rightarrow W^+W^-)$ is model dependent:
The calculation we have presented assumes the minimal Standard Model,
and it will be changed if additional physics beyond this framework is
included.  This effect, however, is most severe far above threshold,
where delicate gauge cancellations are necessary to preserve
unitarity.  Near threshold, the behavior of the cross section is
mainly determined by kinematics, and is less sensitive to the
inclusion of additional physics.  For $\sqrt{s} \approx 160\ {\rm
GeV}$, model dependence is a negligible source of theoretical error.

Uncertainty in the input parameters is also a negligible source of
error.  The input parameters used in calculating the $W^+W^-$ cross
section are the electroweak coupling constants, which, in turn, can be
calculated in terms of $\alpha$, $G_F$, and $M_Z$.  The least well
known of these, $M_Z$, is still known to better than 0.1\%, so this
error is also negligible.

The cross section given in Table~\ref{mass-dependence-table} is
fundamentally a tree-level calculation.  We have included
initial-state radiation, the running of the gauge coupling constants,
and the imaginary part of the $W$'s vacuum polarization, but we have
neglected several other one-loop effects; the dominant such effect is
probably the vacuum polarization of the photon and the $Z$.  By
analogy with physics at the $Z$ peak, we expect that these corrections
are less than 1\%.  Similarly, we expect that the approximations used
in deriving Eq.~(\ref{initial-state-equation}) are of this order.  In
the threshold region, the approximation of Eq.~(\ref{single-smear}) is
better than 1\%, and we neglect it compared to other sources of error.

We will assume, then, that the theoretical error takes the form of an
error in the overall normalization, and that it is approximately
1.5\%.  As discussed in Section~\ref{statistical-section}, this
corresponds to a systematic error in $M_W$ of approximately $20\ {\rm
MeV}$.

\subsection{Systematic error}           \label{systematic-error}
One important source of systematic error is the width of the $W$,
$\Gamma_W$.  In principle, it would be possible to determine both
$\Gamma_W$ and $M_W$ simultaneously, in the fit to $\sigma(s)$.  This
is, however, impractical.  As shown in Fig.~\ref{width-figure}, the
shape of the $W^+W^-$ cross section is not very sensitive to
$\Gamma_W$, particularly in the region where $\sigma(e^+e^-
\rightarrow W^+W^-)$ is most sensitive to $M_W$. We found, by Monte
Carlo simulation, that a simultaneous fit to both $M_W$ and $\Gamma_W$
resulted in a degraded value of $M_W$, while failing to provide a more
precise value for $\Gamma_W$ than that which is already known.

If the Standard Model is assumed to be correct, it is possible to
calculate $\Gamma_W$ with very little theoretical uncertainty.
Unfortunately, we cannot use this calculated value in a measurement of
$M_W$: It is, after all, equally possible to calculate $M_W$ in the
context of the Standard Model, and it would be inconsistent to let
$M_W$ vary while using the predicted value of $\Gamma_W$.  This
inconsistency manifests itself even if $\Gamma_W$ is computed as a
function of $M_W$: Depending on the method of calculation, it scales
either as $M_W^3$ or as $M_W^1$.

We will, then, simply use the measured value of $\Gamma_W$ in our
calculation; the uncertainty in this value will yield a systematic
error in the measurement of $M_W$.  Fortunately, the insensitivity of
the $W^+W^-$ cross section means that this error is small.  Performing
a numerical calculation to be described below, and using 
$\Gamma_W = 2.12 \pm 0.11\ {\rm GeV}$, we obtain $\delta M_W \approx
20\ {\rm MeV}$ 

Other sources of systematic error include uncertainty in the
luminosity and in the calibration of the beam energy.  From
experiences at LEP~100, it is expected that the luminosity will be
known to better than 1\%~\cite{Treille} and that the beam energy will
be known to within 20~MeV.  Uncertainty in the luminosity affects the
overall normalization, while, since we are essentially measuring the
location of the threshold, an error in the energy calibration directly
corresponds to an error in $M_W$.

Finally, performing this measurement requires knowing the efficiency
for detecting and identifying a $W^+W^-$ pair produced nearly at rest.
There are two separate issues: The probability for misidentifying a
$W^+W^-$ event as something else, and the probability for incorrectly
identifying some other type of event as a $W^+W^-$ event.  Events
where at least one $W$ decays to a charged lepton and a neutrino are
probably distinctive enough that misidentification is unlikely, but
events where the final state consists of four jets must be considered
more carefully.

For our purposes, what is important is not the absolute magnitude of
these probabilities, but rather the uncertainty with which they are
known.  This uncertainty, which we denote $\Delta\epsilon$, takes the
form of an additional uncertainty in the overall normalization of the
measured cross section.  The parameter $\Delta\epsilon$ can be
determined only by detailed detector studies, but it is likely that it
will be at most a few percent.  Depending on its magnitude, this could
be the dominant source of systematic error.

We summarize the various sources of systematic and theoretical error
in Table~\ref{systematic-error-table}.

\subsection{Realistic estimate of expected precision}
\label{strategy-section}

We use a fitting procedure to estimate the statistical and systematic
error more precisely: We simulate an experimental run by choosing the
energies at which measurements will be made, and the luminosity that
will be devoted to each measurement.  Using Poisson statistics, we
then randomly generate the number of $W^+W^-$ events observed at each
of these points.  Finally, we perform a numerical fit of the $W^+W^-$
cross section to these randomly generated measurements, taking $M_W$
to be a free parameter; an example of such a fit is shown in
Fig.~\ref{fit-figure}.  We can simulate systematic errors by using a
different value of $\Gamma_W$ or of the normalization constant when
generating the numbers of events than when fitting the cross section.

The systematic and statistical errors are obtained directly by
carrying out this procedure many times: The mean of the difference
between the fitted value of $M_W$ and the true value is the systematic
error, while the variance is the statistical error.  The different
systematic errors are added in quadrature.

We find that, for integrated luminosities between $50\ {\rm pb}^{-1}$
and $1000\ {\rm pb}^{-1}$, and for any reasonable assumptions about
errors in $\Gamma_W$ and in the cross section's normalization, the
optimum strategy is to measure the cross section near threshold, {\it
i.e.}, between 80 and 81~GeV.  Assuming an integrated luminosity of
$100\ {\rm pb}^{-1}$, Fig.~\ref{fit-scan-figure} shows the statistical
error in $M_W$ as a function of the energy at which the cross section
is measured; note that the measurement rapidly becomes ineffective as
the energy is raised much above threshold.

The systematic error due to uncertainty in the luminosity and the
overall normalization of the cross section could be reduced by making
additional measurements further above threshold, but this systematic
error is already less than $50\ {\rm MeV}$, which is sufficiently
small that such an improvement would be negated by the increase in the
statistical error that would result from lowering the statistics in
the region of greatest sensitivity to $M_W$.

We use the values discussed in Sections~\ref{theoretical-error}
and~\ref{systematic-error} to estimate the systematic error.  Adding
the statistical and systematic error in quadrature yields the total
error in measuring $M_W$, as a function of integrated luminosity; this
is plotted in Fig.~\ref{error-figure}.  Until very high luminosities
are obtained, the measurement is dominated by statistics.  Assuming an
integrated luminosity of $100\ {\rm pb}^{-1}$, we estimate that the
total uncertainty will be about $100\ {\rm MeV}$.

\section{Conclusion}
Several methods have been proposed for measuring $M_W$ to higher
precision than is currently available; one of the most promising
methods is to measure the threshold dependence of the total cross
section for $W^+W^-$ pair production at an $e^+e^-$ collider.  

There is, of course, some tension between this measurement and other
physics goals of LEP~200: This measurement requires prolonged running
in an energy region where the $W^+W^-$ cross section is only a few
picobarns, while many other measurements are best made at an energy as
close as possible to the peak $W^+W^-$ cross section.  If it is
possible to obtain an equally good measurement of $M_W$ from
reconstruction of hadronic $W$ decays, that method may, at least
initially, be preferable.

For these purposes, an elaborate calculation of this cross section is
unnecessary: A modified tree level calculation includes most of the
important physical effects, and is sufficiently precise to be compared
to the measured cross section for the extraction of $M_W$.  We have
performed such a calculation, and, using these results, we estimate
that with an integrated luminosity of $100~{\rm pb}^{-1}$ it should be
possible to measure $M_W$ to a precision of about $100\ {\rm MeV}$.


\acknowledgments
This work was supported by the Director, Office of Energy Research,
Office of High Energy and Nuclear Physics, Division of High Energy
Physics of the U.S. Department of Energy under Contract
DE-AC03-76SF00098.


\appendix
\section*{Particles of finite width in the final state}
\label{smear-appendix}
If an unstable particle appears in the final state of a process, the
na{\"\i}ve prescription for the cross section is to calculate the
cross section for the particle to be produced with mass $\sqrt{s}$,
and then convolve this result with some weighting function $\rho(s)$
that resembles a Breit-Wigner.  In this Appendix, we derive a more
precise form of this prescription, making manifest the approximations
that are necessary for the derivation.  We will derive a form for a
final state containing a single $W$; the generalization to two or more
unstable particles is trivial.

The process that actually takes place is the emission of a virtual
$W$, where the final-state particles are leptons and hadrons.  If box
diagrams can be neglected, this process can be factorized into the
production of a $W$ and its decay, {\it i.e.},
\begin{equation}
{\cal M} = P_\mu \Delta^{\mu\nu} {\cal J}_\nu,
\end{equation}
where $P^\mu$ is the part of the amplitude dealing with $W$ production
and with any other particles in the process, $\Delta^{\mu\nu}$ is the
$W$ propagator, and ${\cal J}_\nu$ is the part dealing with $W$ decay.
Since box diagrams are nonresonant, neglecting them is an excellent
approximation.

Phase space factorizes as well: we can write
\begin{equation}
d\Phi = d\Phi(p,q_1,\ldots,q_m) 
        d\Phi_W (k_1, \ldots, k_n)
        (2\pi)^3
        ds,
\end{equation}
where $q_i$ are the momenta of the final-state particles other than
the $W$, $k_i$ are the momenta of the $W$'s decay products, $p$ is the
momentum of the $W$, and $s = p^2$.  The cross section, then, may be
written 
\begin{equation}
d\sigma = A_{\mu\nu} W^{\mu\nu},
\end{equation}
where 
$W^{\mu\nu}$ comes from the $W$ propagator and the $W$'s decay, and 
$A_{\mu\nu}$ is everything else.

By definition, calculating the cross section for production of a $W$
means that we aren't interested in the details of what the $W$ decays
into, so we sum over all decay channels for the $W$ and, for each
channel, integrate over the phase space of the decay products.  We
will deal with a single decay channel; the sum over channels presents
no additional complications.  The result, then, is
\begin{eqnarray}
W^{\mu\nu} &=&				\nonumber
         (2\pi)^3 \int ds
         {\Delta^*}^{\mu\alpha} {\Delta}^{\nu\beta}
         \int d\Phi_W (k_1, \ldots, k_n)
         {{\cal J}^\dagger}_\alpha {{\cal J}}_\beta             \\
    &=&  (2\pi)^3 \int ds
         {\Delta^*}^{\mu\alpha} {\Delta}^{\nu\beta}
         T_{\alpha\beta},               \label{W-tensor}
\end{eqnarray}
where
\begin{equation}
T_{\mu\nu} \equiv 
     \int d\Phi_W (k_1, \ldots, k_n)
          {{\cal J}^\dagger}_\mu {{\cal J}}_\nu.
\end{equation}

It is always possible, of course, to write the integrand in
Eq.~(\ref{W-tensor}) as a product, and thus to write the cross section
in the form
\begin{equation}
\sigma = \int ds \rho(s) \sigma_0(s).
\end{equation}
This statement, by itself, has no physical content; the real content
of Eq.~(\ref{weight-factor}) is that it is possible to define these
functions in such a way that $\rho$ can be interpreted as a
Breit-Wigner and $\sigma_0$ can be interpreted as the cross section
for production of a $W$ with an unphysical value of the mass.  It is
important to remember, however, that $\sigma_0$ is not, in fact, the
cross section for any physical process, and that its meaning must be
defined by explicit construction.

We will define $\sigma_0$ to be the cross section for the production
of a $W$ that has an unphysical mass but that is still on-shell---that
is, that still only has three polarization states.  Consider, then,
the amplitude for the decay of such an on-shell $W$ with momentum $p$
and mass $\sqrt{s}$.  This $W$ is an ordinary vector particle, so the
amplitude is
\begin{equation}
{\cal M} = \epsilon^*_\mu (\sqrt{s}, p, \lambda) {\cal J}^\mu,
\end{equation}
where $\epsilon$ is the $W$'s helicity vector, $\lambda$ is the
helicity index, and $\cal J$ has the same meaning as before.
Averaging over initial helicities, 
\begin{equation}
d\Gamma(s) = 
    {(2\pi)^4 \over 2\sqrt{s}} d\Phi_W          \cdot
    {1\over3} {1\over s}
    (p^\mu p^\nu - s g^{\mu\nu}) 
    {{\cal J}^\dagger}_\mu {{\cal J}}_\nu,
\end{equation}
where we have used the fact that
\begin{equation}
\sum_\lambda \epsilon^*_\mu (\sqrt{s}, p, \lambda)
             \epsilon  _\nu (\sqrt{s}, p, \lambda) 
   = {1\over s} (p_\mu p_\nu - s g_{\mu\nu}).
\end{equation}
Performing the phase space integral, 
\begin{equation}        \label{w-decay-rate}
\Gamma(s) = 
    {(2\pi)^4 \over 2\sqrt{s}} 
    {1\over3} {1\over s}
    (p^\mu p^\nu - s g^{\mu\nu}) 
    T_{\mu\nu}.
\end{equation}

If the masses of all fermions in the final state may be neglected,
then the $W$ couples to a conserved current, and we may write
\begin{equation}        \label{T-lorentz-form}
T^{\mu\nu} = (p^\mu p^\nu - s g^{\mu\nu}) T(s).
\end{equation}
In the case at hand this approximation is permissible: The heaviest
fermion that a $W$ can decay to is the $b$ quark, and
$m_b^2 / M_W^2 < 0.004$.  Eq.~(\ref{w-decay-rate}) thus simplifies to
\begin{equation}        \label{simple-w-decay-rate}
\Gamma(s) = (2\pi)^4 \sqrt{s} {T(s) \over 2}.
\end{equation}
Note, however, that this simplifying approximation is crucial: If the
masses of particles in the final states cannot be neglected, then no
such simple form as Eq.~(\ref{weight-factor}) exists.

Similarly, returning to $W^{\mu\nu}$, Eq.~(\ref{W-tensor}) simplifies
as well.  The gauge-dependent piece of the $W$ propagator vanishes
when contracted with a tensor of the form given in
Eq.~(\ref{T-lorentz-form}), so we may write
\begin{equation}
\Delta^{\mu\nu} = \Delta(s) g^{\mu\nu},
\end{equation}
and, using Eq.~(\ref{simple-w-decay-rate}), 
\begin{equation}
W_{\mu\nu} = 
\int ds \left|\Delta(s) \right|^2
        {\sqrt{s} \Gamma(s) \over\pi}
        \sum_\lambda \epsilon^*_\mu (\sqrt{s}, p, \lambda)
                     \epsilon  _\nu (\sqrt{s}, p, \lambda).
\end{equation}
The $W_{\mu\nu}$ tensor is the only part of the cross section that
depends on the details of the $W$ in the final state, and this sum
over the product of polarization vectors is the form that $W_{\mu\nu}$
would have if the final-state particle were a $W$ of mass $\sqrt{s}$, 
so we have established that
\begin{equation}                \label{penultimate-smear}
\sigma = \int ds \left|\Delta(s) \right|^2
                 {\sqrt{s} \Gamma(s) \over\pi}
                 \sigma_0 (s).
\end{equation}

Now consider the form of $\Delta(s)$.  At tree level, 
\begin{equation}
\Delta(s) = {-i \over s - M_W^2 + i\epsilon},
\end{equation}
but, including the vacuum polarization of the $W$, it becomes
\begin{equation}
\Delta(s) = {-i \over s - M_W^2 + \delta M_W^2 - \Pi_{WW}(s)}.
\end{equation}
If $M_W$ is taken to be the physical $W$ mass, then 
${\rm Re} \: \Pi_{WW}(M_W^2)$ exactly cancels the $W$ mass counterterm
$\delta M_W^2$, so
\begin{equation}
\Delta(s) = {-i \over s - M_W^2 - i {\rm Im} \: \Pi_{WW}(s) 
                                - {\rm Re} \: \widetilde\Pi_{WW}(s)},
\end{equation}
where $\widetilde\Pi(s) \equiv \Pi(s) - \Pi(M_W^2)$.  We are only
interested in $s \approx M_W^2$, and ${\rm Re} \: \widetilde\Pi(s)$ is
a smooth function of $s$ that, by construction, vanishes at $s =
M_W^2$.  It can thus be absorbed into the wave-function
renormalization, and we neglect it as a non-leading correction.

Using the optical theorem, we can show~\cite{cahn} that 
\begin{equation}
- {\rm Im} \: \Pi_{WW}(s) = \sqrt{s} \Gamma(s),
\end{equation}
where $\Gamma(s)$ is the decay rate for an on-shell $W$ of mass
$\sqrt{s}$.  Finally, then, we can evaluate $\Gamma(s)$ either by
performing the tree-level calculation, or by scaling the measured
value.  We choose the latter method.  If the $W$'s decay products are
massless, then $\Gamma(s)$ scales as $\sqrt{s}$, so
\begin{equation}
\Gamma(s) = \sqrt{s} {\Gamma \over M_W},
\end{equation}
where $\Gamma$ is the measured width of the physical $W$.

Substituting into Eq.~(\ref{penultimate-smear}), we obtain
\begin{eqnarray}
\sigma &=& \int ds 
                {1 \over \left(s - M_W^2 \right)^2 + s \Gamma^2(s)}
                {\sqrt{s} \Gamma(s) \over\pi}
                \sigma_0 (s)                            \\
       &=& \int ds 
                {1 \over \left(s - M_W^2 \right)^2 
                        + s^2 \Gamma^2 / M_W^2}
                {s \Gamma \over \pi M_W}
                \sigma_0 (s),
\end{eqnarray}
thus verifying Eq.~(\ref{weight-factor}).



\mediumtext
\begin{table}
\squeezetable
\caption{
$W^+W^-$ pair production cross section, including the finite value of
$\Gamma_W$ and the effects of initial-state radiation, for seven
values of the $W$ mass.  The present value is $M_W = 80.22 \pm 0.26\ 
{\rm GeV}$.  The cross section is a smooth function of $M_W$, and
linear interpolation should suffice for values of $M_W$ in between
those presented here.  
}
\label{mass-dependence-table}
\begin{tabular}{lddddddd}
$\sqrt{s}$ (GeV) & \multicolumn{7}{c}{$\sigma\ ({\rm pb})$} \\ 
& \multicolumn{1}{r}{79.6 GeV}
& \multicolumn{1}{r}{79.8 GeV}
& \multicolumn{1}{r}{80.0 GeV}
& \multicolumn{1}{r}{80.2 GeV}
& \multicolumn{1}{r}{80.4 GeV}
& \multicolumn{1}{r}{80.6 GeV}
& \multicolumn{1}{r}{80.8 GeV}
\\ \tableline
150 &0.447 &0.424 &0.402 &0.382 &0.363 &0.346 &0.329\\ 
151 &0.508 &0.480 &0.454 &0.430 &0.408 &0.387 &0.368\\ 
152 &0.582 &0.547 &0.516 &0.487 &0.461 &0.436 &0.413\\ 
153 &0.675 &0.632 &0.593 &0.557 &0.525 &0.495 &0.468\\ 
154 &0.793 &0.738 &0.689 &0.644 &0.604 &0.567 &0.534\\ 
155 &0.951 &0.877 &0.812 &0.755 &0.704 &0.657 &0.616\\ 
156 &1.169 &1.067 &0.978 &0.901 &0.832 &0.772 &0.719\\ 
157 &1.485 &1.336 &1.209 &1.100 &1.007 &0.925 &0.854\\ 
158 &1.965 &1.738 &1.547 &1.387 &1.252 &1.137 &1.038\\ 
159 &2.689 &2.353 &2.064 &1.820 &1.615 &1.443 &1.298\\ 
160 &3.651 &3.225 &2.831 &2.478 &2.170 &1.908 &1.688\\ 
161 &4.704 &4.256 &3.812 &3.383 &2.978 &2.609 &2.284\\ 
162 &5.728 &5.293 &4.853 &4.410 &3.971 &3.540 &3.129\\ 
163 &6.676 &6.268 &5.851 &5.427 &4.996 &4.561 &4.128\\ 
164 &7.539 &7.159 &6.771 &6.373 &5.968 &5.554 &5.133\\ 
165 &8.324 &7.970 &7.608 &7.238 &6.860 &6.473 &6.078\\ 
166 &9.038 &8.707 &8.369 &8.024 &7.672 &7.311 &6.943\\ 
167 &9.689 &9.378 &9.061 &8.744 &8.409 &8.073 &7.730\\ 
168 &10.284 &9.992 &9.694 &9.391 &9.082 &8.767 &8.446\\ 
169 &10.830 &10.552 &10.272 &9.987 &9.696 &9.400 &9.099\\ 
170 &11.330 &11.068 &10.803 &10.533 &10.257 &9.979 &9.696\\ 
171 &11.789 &11.541 &11.289 &11.034 &10.774 &10.510 &10.241\\ 
172 &12.212 &11.976 &11.736 &11.494 &11.247 &10.997 &10.744\\ 
173 &12.601 &12.376 &12.148 &11.917 &11.683 &11.445 &11.204\\ 
174 &12.961 &12.746 &12.528 &12.307 &12.084 &11.858 &11.629\\ 
175 &13.294 &13.088 &12.878 &12.668 &12.454 &12.238 &12.020\\ 
176 &13.601 &13.403 &13.203 &13.001 &12.797 &12.590 &12.381\\ 
177 &13.885 &13.695 &13.503 &13.309 &13.113 &12.915 &12.715\\ 
178 &14.148 &13.965 &13.781 &13.595 &13.406 &13.216 &13.024\\ 
179 &14.391 &14.215 &14.037 &13.858 &13.678 &13.495 &13.310\\ 
180 &14.616 &14.446 &14.275 &14.102 &13.928 &13.752 &13.575\\ 
181 &14.824 &14.660 &14.495 &14.328 &14.160 &13.991 &13.820\\ 
182 &15.016 &14.857 &14.698 &14.537 &14.375 &14.212 &14.047\\ 
183 &15.193 &15.040 &14.886 &14.730 &14.574 &14.416 &14.257\\ 
184 &15.357 &15.209 &15.060 &14.909 &14.758 &14.605 &14.451\\ 
185 &15.508 &15.365 &15.220 &15.075 &14.928 &14.780 &14.631\\ 
186 &15.648 &15.508 &15.368 &15.227 &15.085 &14.942 &14.798\\ 
187 &15.776 &15.641 &15.505 &15.368 &15.230 &15.091 &14.952\\ 
188 &15.894 &15.767 &15.630 &15.497 &15.364 &15.229 &15.093\\ 
189 &16.002 &15.875 &15.746 &15.621 &15.487 &15.356 &15.225\\ 
190 &16.102 &15.977 &15.852 &15.727 &15.600 &15.477 &15.345\\ 
\end{tabular}
\end{table}
\narrowtext

\mediumtext
\begin{table}
\caption{
Sources of systematic error in the measurement of $M_W$ by study of
the threshold, as discussed in Sections~\protect\ref{systematic-error}
and~\protect\ref{strategy-section}; errors are added in
quadrature.  Many of these errors take the form of uncertainty in the
overall normalization, which allows us to use
Eq.~(\protect\ref{norm-error-equation}).  We are unable to estimate the
uncertainty in detection efficiency, so we parameterize it as
$\Delta\epsilon$.  In our plots, we assume $\Delta\epsilon = 2\%$.
All of these errors are sufficiently small that the measurement will
be dominated by statistical error.  }
\label{systematic-error-table}
\begin{tabular}{lcc}
Effect  &  Effect on normalization      & Error in $M_W$        \\
\tableline              
Coupling constants      & negligible    & negligible            \\
Model dependence        & negligible    & negligible            \\
One-loop corrections    & 1\%           & 17 MeV                \\
Initial-state radiation & 1\%           & 17 MeV                \\
Smearing approximation  & negligible    & negligible            \\
Total theoretical error & 1.5\%         & 20 MeV                \\
                                \tableline
$\Gamma_W$              &               & 20 MeV                \\
Luminosity              & 1\%           & 17 MeV                \\
Beam energy             &               & 20 MeV                \\
Detection efficiency    & $\Delta\epsilon$
                        & $(17\ {\rm MeV}) \times \Delta\epsilon$  \\
\tableline
Total systematic error  &               & 45 MeV                 \\
\end{tabular}
\end{table}
\narrowtext

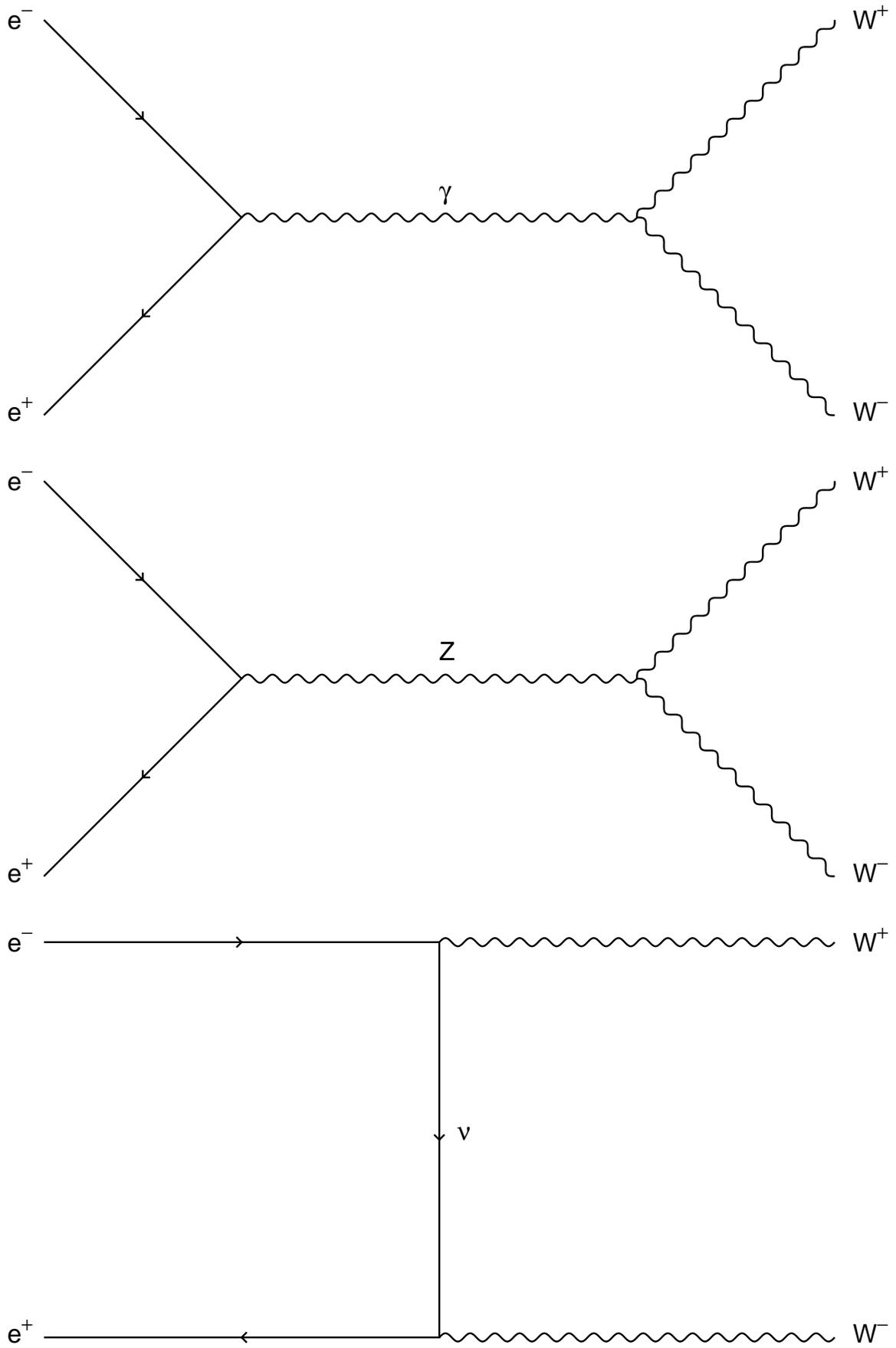
\begin{figure}
\caption{
Tree-level Feynman diagrams for $W^+W^-$ pair production at an
$e^+e^-$ collider.  Exchange of a virtual Higgs boson also
contributes, but is negligible.
}
\label{production-diagrams}
\end{figure}
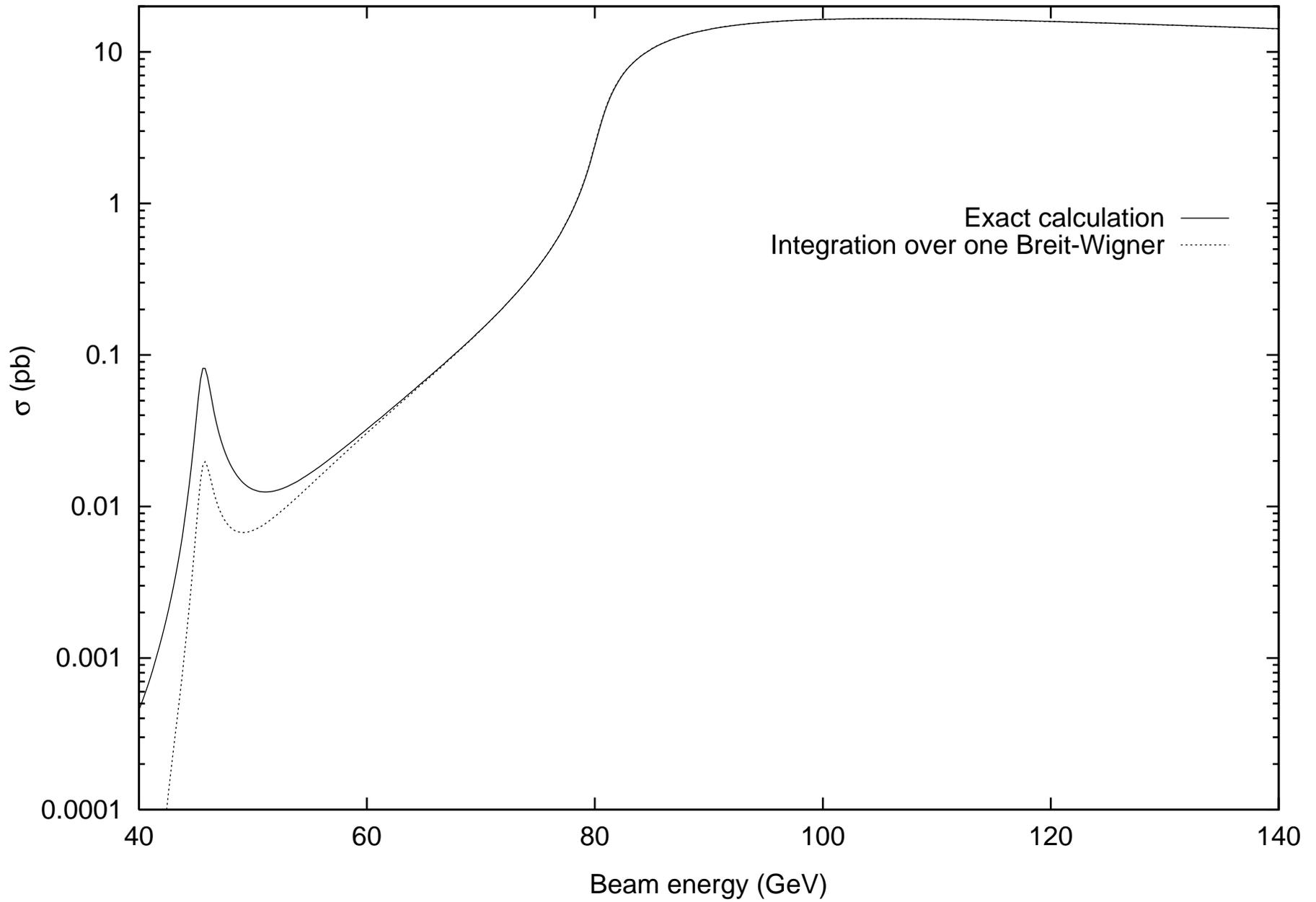
\begin{figure}
\caption{
Comparison of an exact tree-level calculation of $\sigma(e^+e^-
\rightarrow W^+W^-)$, in which both $W$'s in the final state are
integrated over Breit-Wigners, to an approximate calculation, in which
only one $W$ is integrated over a Breit-Wigner, but using a width
twice the physical value.  Note that this is an excellent
approximation except in a domain so far below threshold that the cross
section is unobservably small.  The peak at $45\ {\rm GeV}$ is due to
the decay of a $Z$ to two virtual $W$'s.
}
\label{single-smear-comparison}
\end{figure}
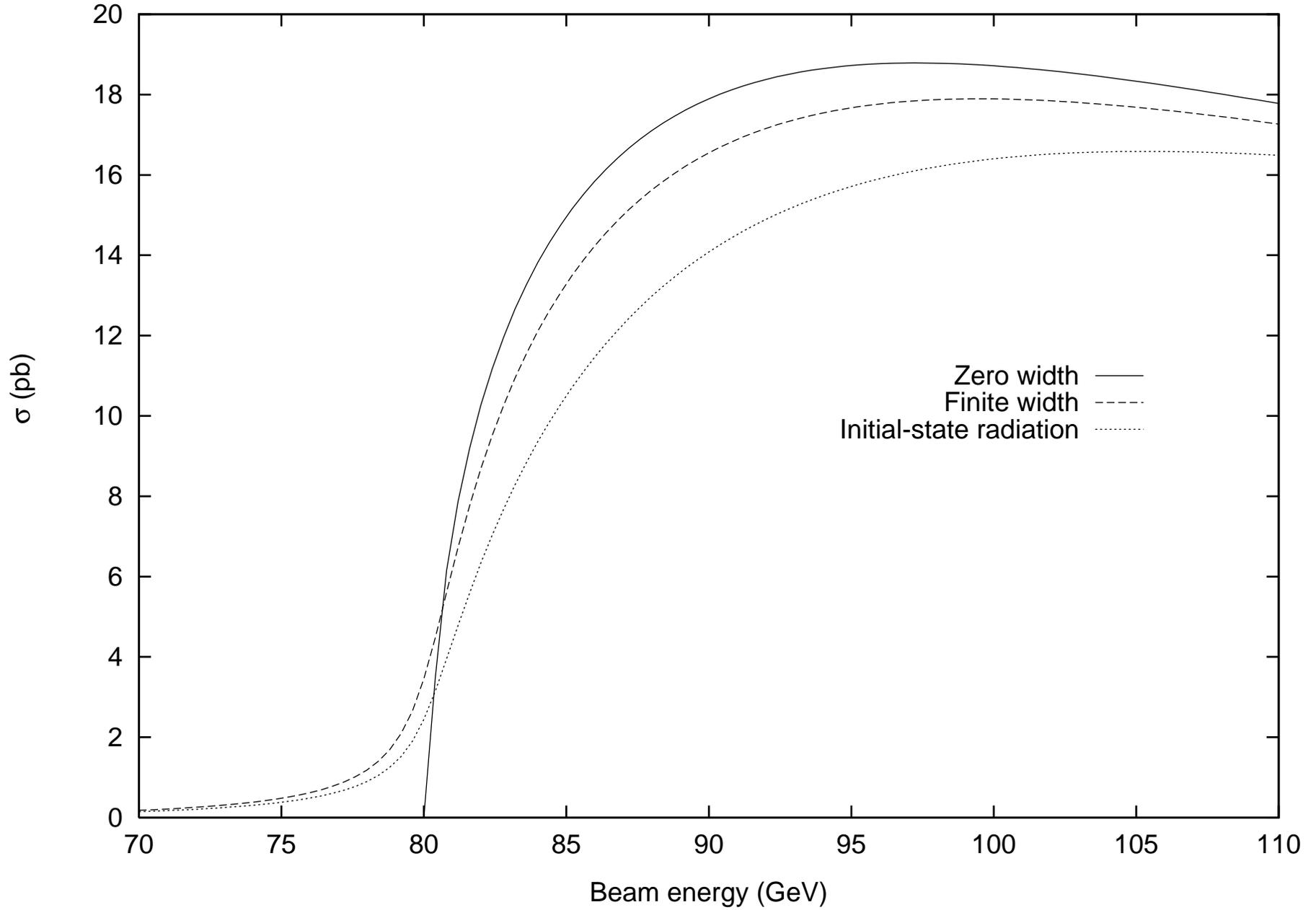
\begin{figure}
\caption{
Calculation of $\sigma(e^+e^- \rightarrow W^+W^-)$.  The solid curve
is the tree-level cross section where both final-state $W$'s are
on-shell, the dashed curve is the cross section where the $W$'s are
allowed to be off-shell, and the dotted curve includes initial-state
radiation as well as the finite $W$ width.  Note that the finite width
of the $W$ and the initial-state radiation are effects of comparable
magnitude.
}
\label{cross-section-plot}
\end{figure}
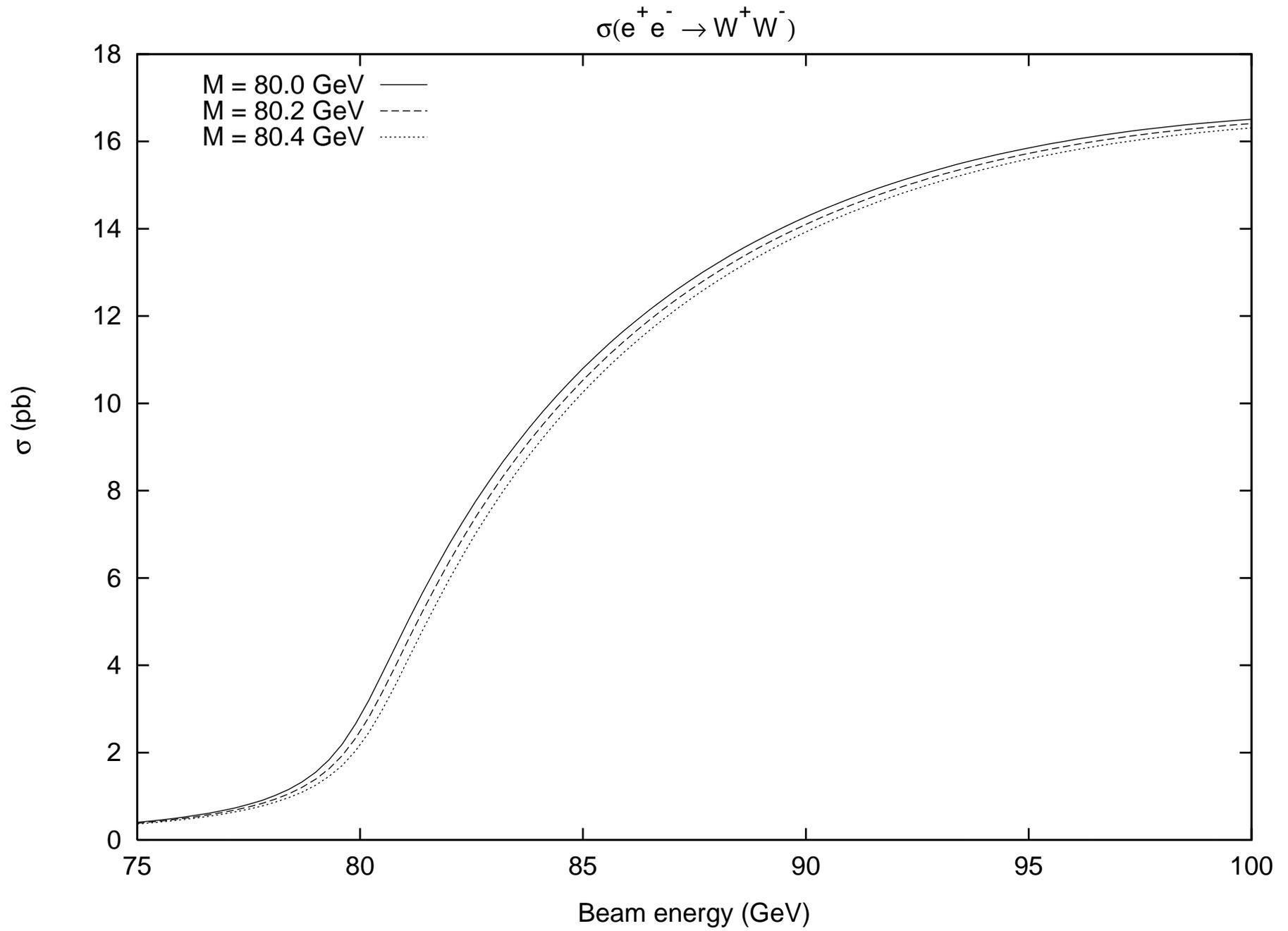
\begin{figure}
\caption{
Calculation of $\sigma(e^+e^- \rightarrow W^+W^-)$, including the
finite width of the $W$ and the effect of initial-state radiation.
The solid curve is for $M_W = 80.0\ {\rm GeV}$, the dashed curve is
for $M_W = 80.2\ {\rm GeV}$, and the dotted curve is for $M_W = 80.4\
{\rm GeV}$.  The dependence on $M_W$ is strongest for $E \approx M_W$.
}
\label{several-masses}
\end{figure}
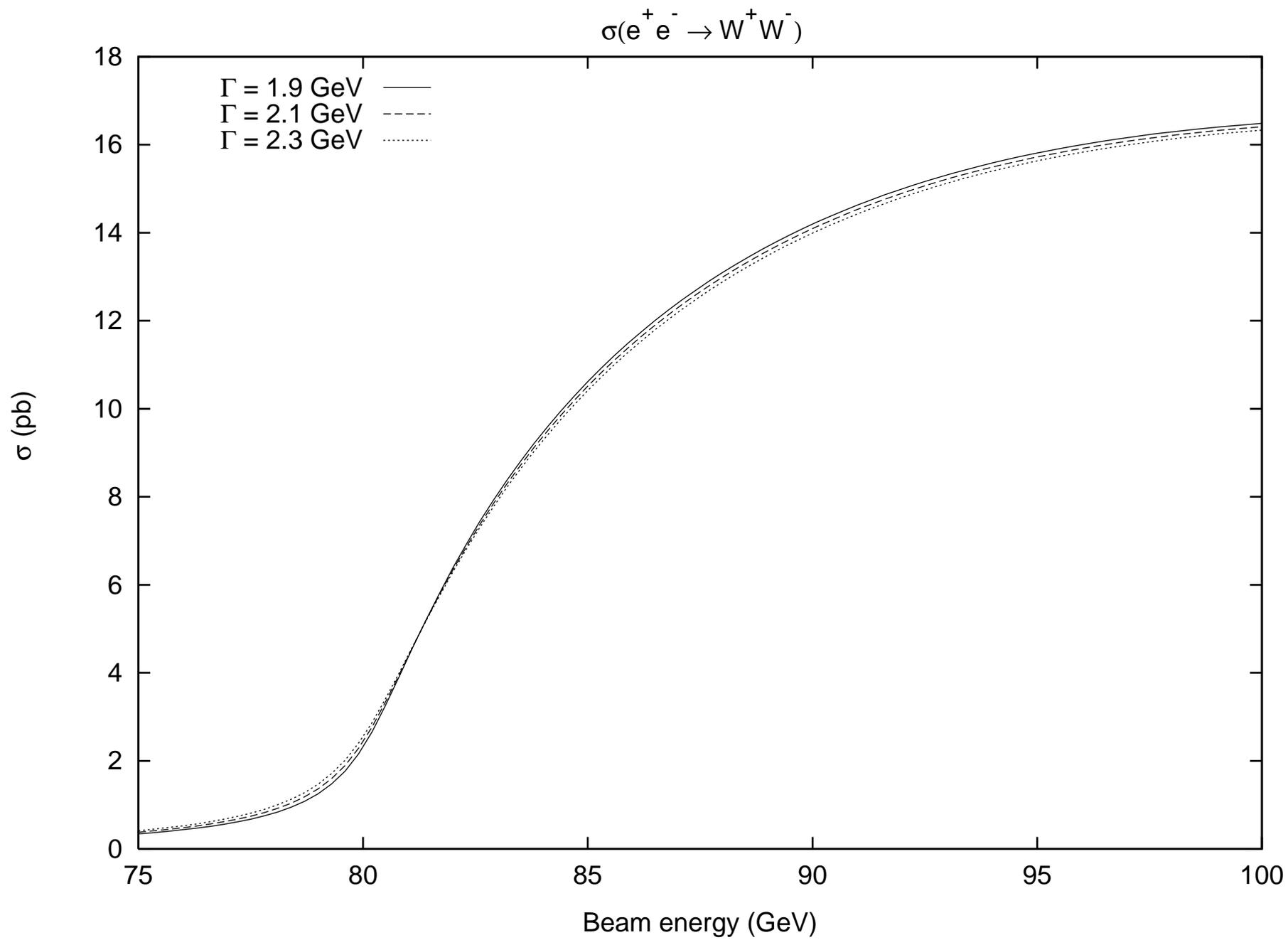
\begin{figure}
\caption{
Calculation of $\sigma(e^+e^- \rightarrow W^+W^-)$, including the
finite width of the $W$ and the effect of initial-state radiation.
The three curves represent different values for $\Gamma_W$: the solid
curve is with $\Gamma_W = 1.9\ {\rm GeV}$, the dashed curve is with
$\Gamma_W = 2.1\ {\rm GeV}$, and the dotted curve is with $\Gamma_W =
2.3\ {\rm GeV}$.  The measured value is $\Gamma_W = 2.12 \pm 0.11\
{\rm GeV}$.  Note that for values of the width close to the measured
value, the form of the cross section is not very sensitive to
$\Gamma_W$.
}
\label{width-figure}
\end{figure}
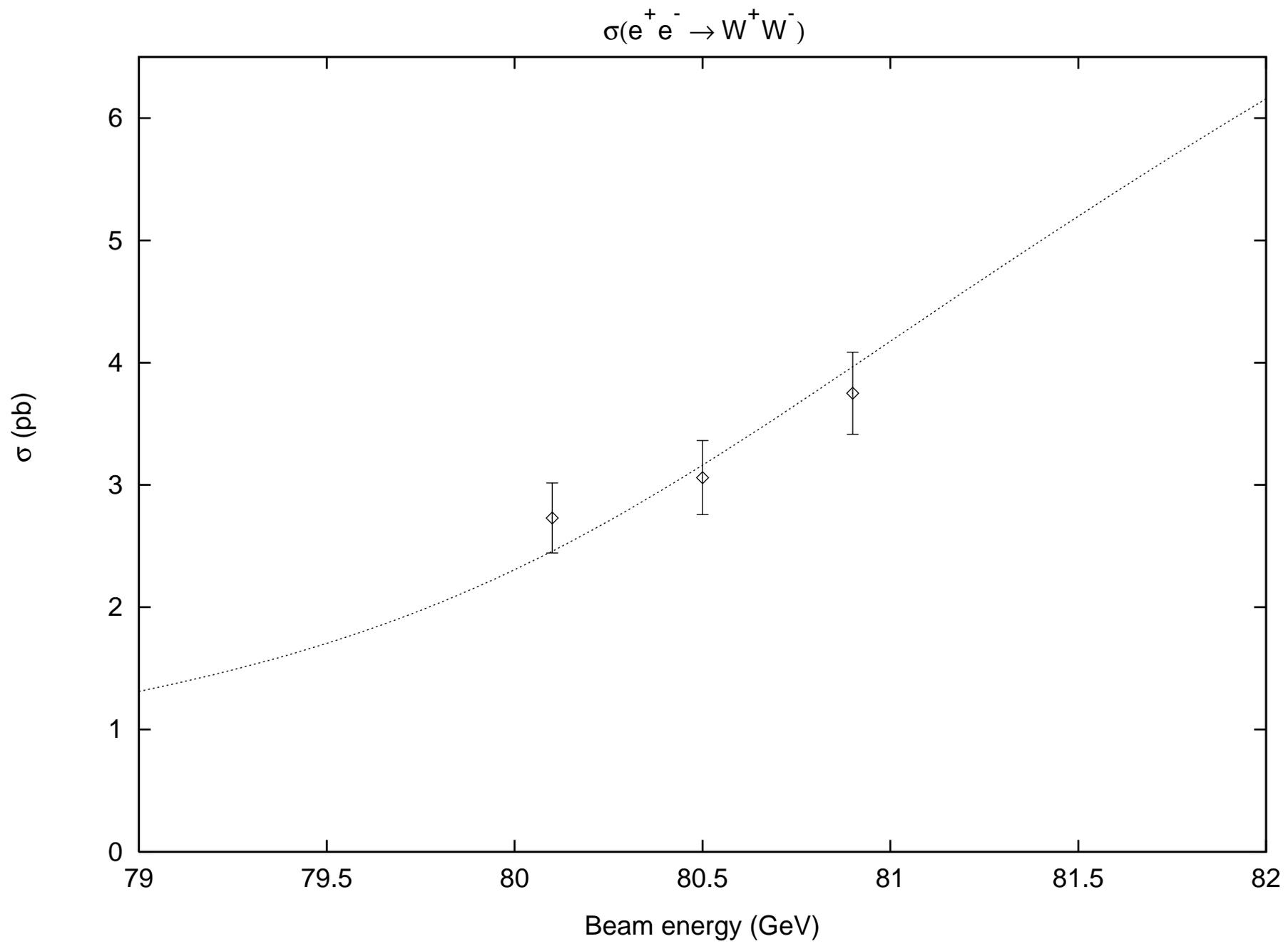
\begin{figure}
\caption{
Simulation of an experimental determination of $M_W$, using an
integrated luminosity of $100 {\rm pb}^{-1}$.  Measurements are made
at $80.1\ {\rm GeV}$, $80.5\ {\rm GeV}$, and $80.9\ {\rm GeV}$.  At
each point, the number of observed events is randomly chosen, using a
Poisson distribution whose mean is the cross section times the
integrated luminosity.  The points, with 1-$\sigma$ error bars,
represent these simulated measurements of $\sigma$ at those three
energies, and the solid curve, which corresponds to $M_W = 80.31\ {\rm
GeV}$, is the best fit.  
}
\label{fit-figure}
\end{figure}
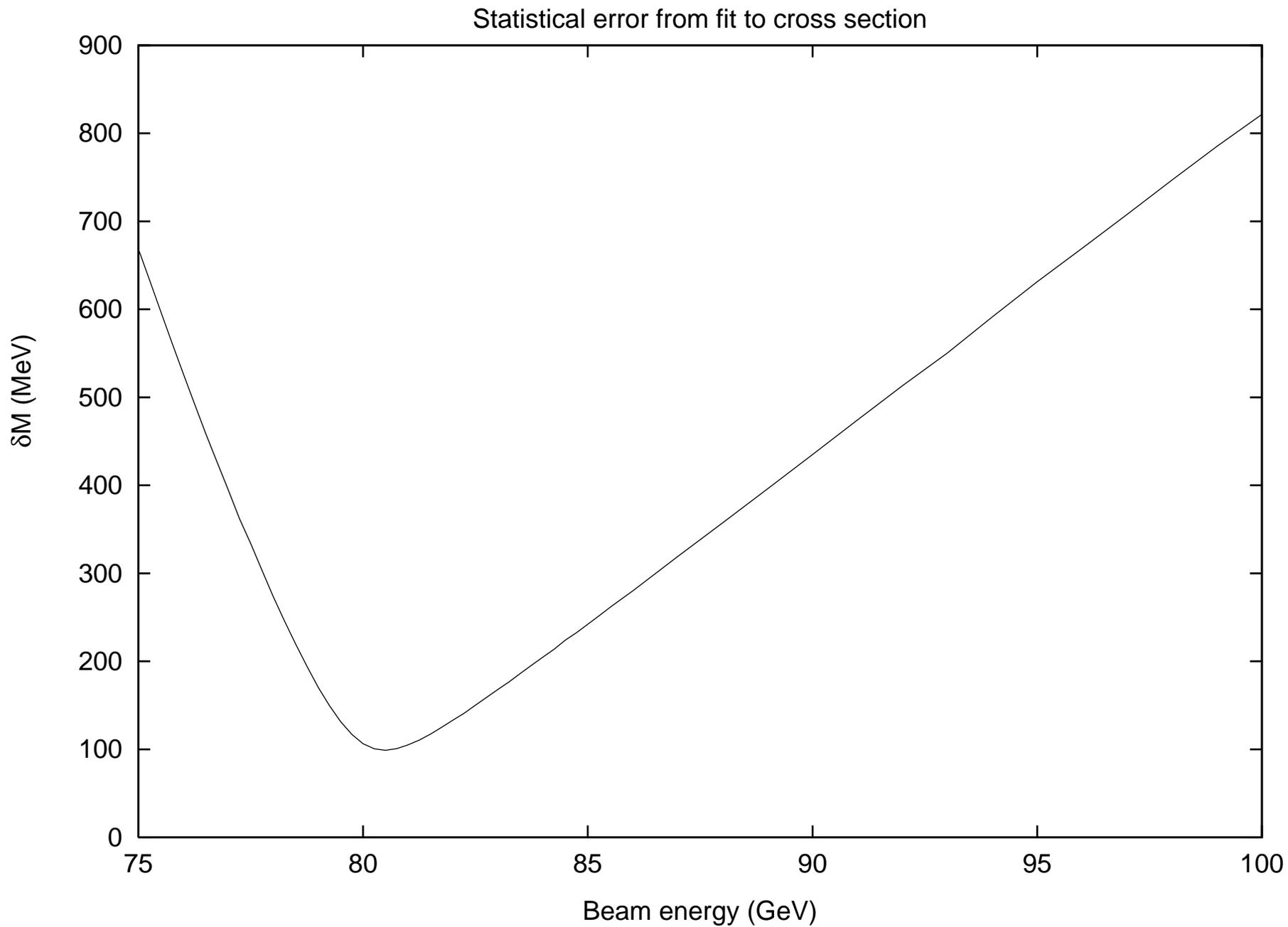
\begin{figure}
\caption{
Statistical error in the measurement of $M_W$, in MeV, as a function
of the energy at which $\sigma(e^+e^- \rightarrow W^+W^-)$ is measured.
Note that the statistical error is quite high unless the measurement
is made at the $W^+W^-$ threshold.  This plot is generated assuming an
integrated luminosity of $100\ {\rm pb}^{-1}$, but this qualitative
feature is true independent of luminosity.  The total cross section at
this energy is roughly $0.3\ {\rm pb}$, compared to a peak value of
more than $16\ {\rm pb}$.
}
\label{fit-scan-figure}
\end{figure}
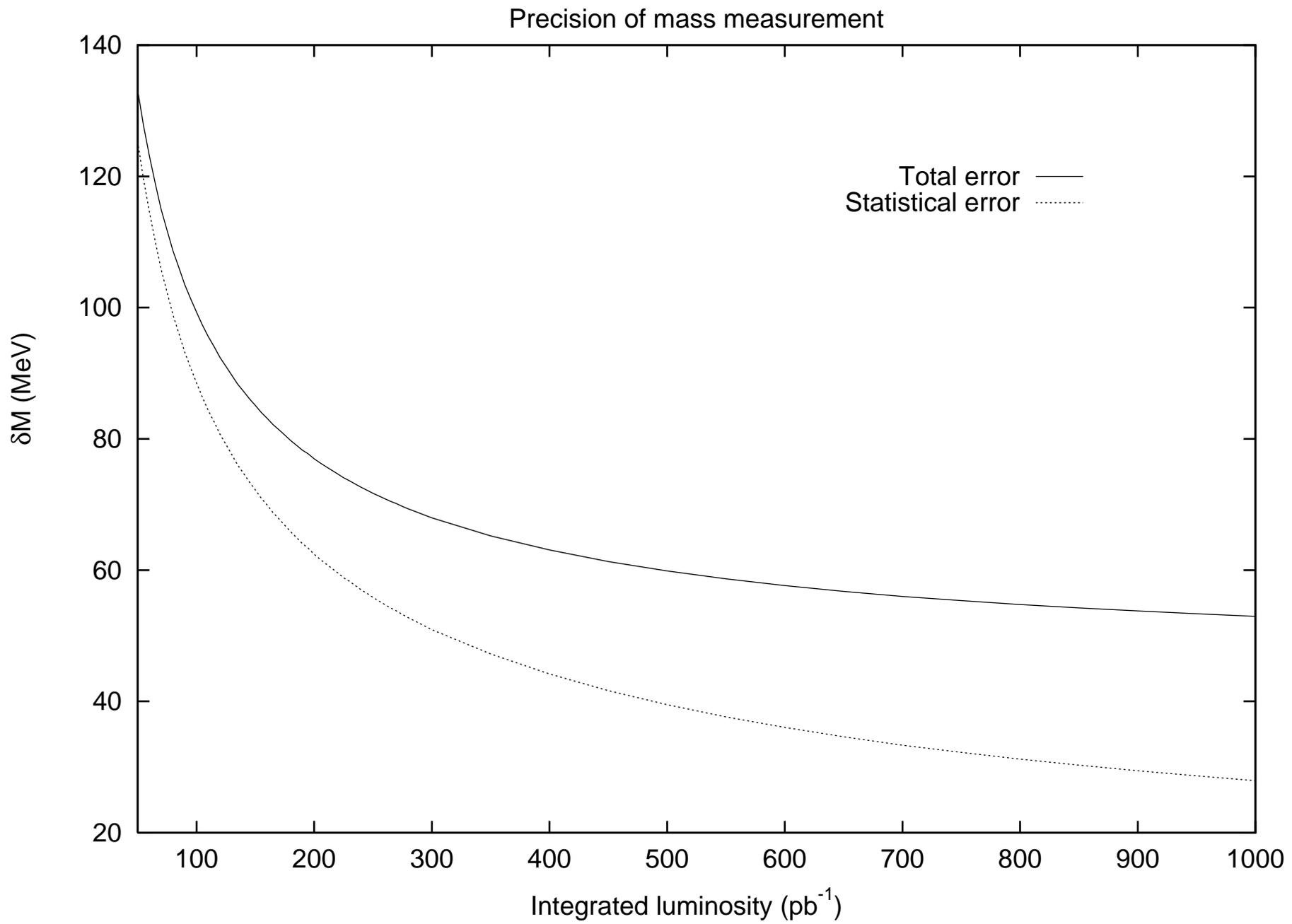
\begin{figure}
\caption{
Estimated error in the measurement of $M_W$, in MeV, as a function of
integrated luminosity.  The solid curve is the total error, and the
dotted curve is the statistical error.  Note that until very high
luminosities are obtained, the measurement is dominated by statistical
error.  
}
\label{error-figure}
\end{figure}
\end{document}